\documentstyle[twocolumn,aps]{revtex}
\begin{document}
\input{psbox}
\draft

\title{
Magnetoresistance in La$_{1-x}$ Sr$_x$CoO$_3$ for 0.05$\leq x \leq$0.25
}

\author{Vladimir Golovanov and Laszlo Mihaly}
\address{Department of Physics, State University of New York at Stony Brook, 
Stony Brook, NY 11794-3800}
\author{A. R. Moodenbaugh}
\address{Department of Applied Science, Brookhaven National Laboratory, Upton,
NY 11973-5000}

\date{\today}
\maketitle

\begin{abstract}
The dc resistivity, magnetoresistance and magnetic susceptibility of 
La$_{1-x}$Sr$_x$CoO$_3$ compounds have been investigated in the temperature 
range of 4K to 300K for magnetic fields up to 7 T.  In the doping 
range studied (0.05$\leq x \leq$0.25) the electronic properties of the 
material exhibit a crossover from semiconducting to metallic behavior.
The magnetoresistance is highest in the semiconducting 
state.  A correlation was found between the energy gap determined from 
the dc conductivity and the energy scale identified from 
neutron scattering data.  The results are interpreted in terms of a 
double exchange model.  
\end{abstract}
\pacs{PACS: 72.15.Gd, 72.20.My, 71.45.Gm}

\narrowtext
The recent discovery of colossal magnetoresistance (MR) in thin films of La-
Ca-Mn-O \cite{jin}, \cite{mccormack} and giant magnetoresistance in a 
ferromagnetic perovskite of La-Ba-Mn-O \cite{helmholt} generated a 
renewed interest in this family of compounds.  
Magnetization and resistivity studies of La$_{1-x}$Sr$_x$MnO$_3$ single crystals 
\cite{urushibara} revealed several phases with the
the highest magnetoresistance observed at the paramagnetic insulator 
to ferromagnetic metal transition. 
Neutron scattering measurements on La$_{0.7}$Sr$_{0.3}$MnO$_3$
\cite{martin} demonstrated that the ferromagnetism in 
La$_{0.7}$Sr$_{0.3}$MnO$_3$ is itinerant in character.

Although most of the recent attention has been focused on the MnO$_3$ 
perovskites, similar properties has been observed in materials based on 
CoO$_3$. The first studies of magnetic and transport properties of 
La$_{1-x}$Sr$_x$CoO$_3$ by Jonker and van Santen  \cite{jonker} were 
interpreted by Goodenough \cite{goodenough}. 
Recently Se\~nar\'{\i}s-Rodr\'{\i}guez and Goodenough
performed extensive magnetic and transport 
studies of pure LaCoO$_3$ \cite{goodenough3} and doped 
La$_{1-x}$Sr$_x$CoO$_3$ \cite{goodenough2}.
Itoh {\it et al.} \cite{itoh} deduced the magnetic phase diagram of 
La$_{1-x}$Sr$_x$CoO$_3$ from magnetization measurements.
Three phases were identified:    
at low temperatures spin-glass (for $x<0.18$) and cluster-glass 
(for $x>0.18$) phases and, at high temperatures, a paramagnetic phase.
The magnetization dependence of the resistivity of La$_{1-x}$Sr$_x$CoO$_3$ 
single crystals was investigated for $x>0.2$ by Yamaguchi 
{\it et al.} \cite{yamaguchi}.  
The electronic structure of the material was studied near the 
semiconductor-metal transition in La$_{1-x}$Sr$_x$CoO$_3$ by using 
electron-spectroscopy \cite{chainani}.   

The negative magnetoresistance in the transition metal perovskites is 
usually interpreted in terms of the ``double exchange" mechanism, 
suggested by Zener \cite{zener}, and developed by Anderson 
\cite{anderson} and DeGennes \cite{degennes2}.  
The principal idea is that most of the electrons on the outer shells of 
the transition metal reside on localized orbits, 
coupled by Hund's rule to large magnetic moments, 
whereas others participate in the conduction {\it via} 
overlapping orbits.  Due to the exchange interaction between the two 
types of electrons, the conduction is conditional on the appropriate 
orientation of the underlying localized moments.  
A related approach, suggested for metals by DeGennes and Friedel 
\cite{degennes1} and adapted to semiconductors by Haas {\it et al.} 
\cite{haas}
treats the magnetic moments in a mean field approximation.  The 
``perfect" ferromagnetic order leads to a spin splitting of the 
conduction band, whereas the magnetic disorder is viewed as a source for 
extra scattering processes.  

The magnetoresistance, and the electrical conduction in general, is 
strongly influenced by the spin state of the Co ions, which was the subject 
of recent neutron scattering measurements by Asai {\it et al.} \cite{asai}. 
Motivated by this study, we investigated the low and high field magnetization, dc 
electrical resistivity and the magnetoresistance, for magnetic fields 
up to 7 T, on a set of ceramic samples of composition 
La$_{1-x}$Sr$_x$CoO$_3$.  In contrast to the work by Yamaguchi {\it et al.}
\cite{yamaguchi}, we concentrated on the low doping range, 
$0.05\leq x \leq0.25$.    


La$_{1-x}$Sr$_x$CoO$_3$ polycrystalline samples were prepared by solid state  
reaction method similar to that described in \cite{itoh}.
The appropriate mixture of La$_2$O$_3$, SrCO$_3$ and CoO was 
ground and calcined repeatedly at $950^{\circ}$C for 10 days, fired 
at $1300^{\circ}$C   for 
about 28 hours and then cooled in air
at a rate of approximately $100^{\circ}$C/hour.  This cooling rate is 
considered to be ``fast", as opposed to ``slow" cooling rate of $100^{\circ}$C/day 
used by Itoh {\em et al.} \cite{itoh}.  Fast cooling ($60^{\circ}$C/hour) has been 
also used in the recent work of Se\~nar\'{\i}s-Rodr\'{\i}guez and Goodenough
\cite{goodenough2}.
The samples were confirmed to be of a single phase 
with rhombohedrally distorted perovskite structure  by powder X-ray diffraction
analysis.
The low field magnetic properties of the samples produced here agreed well with the 
published results \cite{itoh} and the resistivity curves for $x$=0.2 and 0.25 
were similar to those obtained in Ref.\cite{goodenough2}.
We found, however, that the temperature dependent resistivity 
of different cuts from the same specimen were different. 
In order to remedy this shortcoming, we 
performed an additional heat treatment at $950^{\circ}$C for 5 hours and 
we cooled the samples slowly, at a rate of $100^{\circ}$C per day as suggested by 
Itoh {\it et al.} \cite{itoh}.  After the heat treatment the low field magnetic
properties did not change significantly, but the resistivity did: In contrast to
the fast cooled specimens \cite{goodenough2} the resistivity curves  
for $x$=0.2 and 0.25 had positive slope for the whole temperature  range
measured. The resistivity measurements were very reproducible for all 
compositions. The data reported here were obtained on the slow cooled 
material.  

\begin{figure}
\pswdincr=-3cm
\psboxto(3.4in;0in){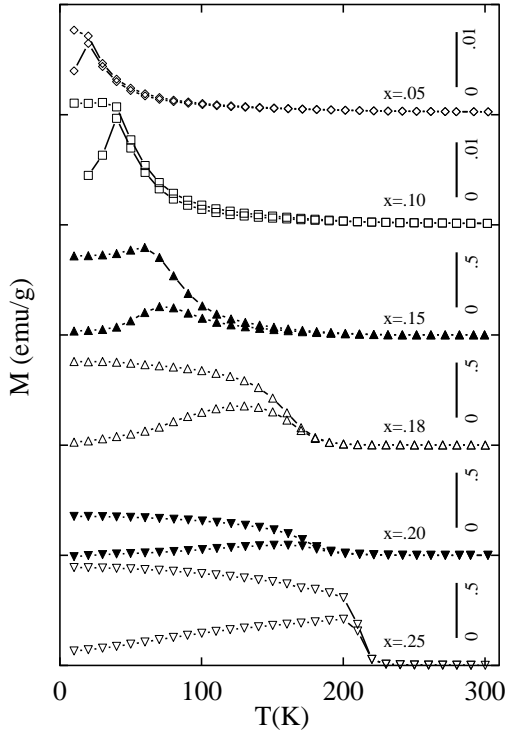}
\caption{Low field magnetization measurements on samples of 
various Sr content $x$.  For each $x$, the lower magnetization was 
obtained in the zero field
cooled measurement, the higher one corresponds to field cooling.  
The curves are shifted for better view; the 
high temperature magnetization is close to zero for each samples.  The 
vertical bars indicate the magnetization scale.  For $x=0.05$ and 
$x=0.1$ the scale is expanded by a factor of 50.}
\label{magneti}
\end{figure}

Magnetization measurements were performed
using a SQUID magnetometer in low (20 Gauss) and high (50 kGauss = 5T) magnetic 
fields. For the low field measurements the samples were first cooled 
in zero magnetic field 
(ZFC measurement),  then in the 20G magnetic field (FC measurement).  
A difference between the ZFC and FC data indicates a magnetic phase with
permanent magnetization and hysteresis in the magnetization curve.   
To estimate the saturation magnetization of the system 
we took ZFC data for the high field.

The electrical resistance was measured as a function of temperature and  
magnetic field in a superconducting magnet with
the maximum applied field of $H$=7T. 
The samples were rectangular in shape and about 10x5x2 mm in size. 
Four electrical leads were glued with silver paste, 
in line, along the long axis of the specimen.
The outside leads were used to supply the current. 
The voltage drop was measured on the inside leads.  
The direction of current was perpendicular to magnetic field.
The linearity in the current-voltage 
dependence has been checked at several temperatures and magnetic fields;  
for the range of currents used here 
the resistivity proved to be ohmic for all samples. 

To investigate the magnetoresistance we swept the magnetic field at 
several fixed temperatures.  This method gives a high accuracy (especially  
for semiconductor samples, where a temperature lag between the sample and the 
thermometer could easily lead to a resistance difference larger than the 
magnetoresistance), and it also provides a full picture of the possible
non-linearity and hysteresis of the MR.   

Magnetization measurements on our samples (Fig. \ref{magneti}) led to results 
similar to those observed by Itoh et al. \cite{itoh}.
At the higher $x$ values the samples exhibit ferromagnetism, with a 
Curie temperature of 220K for $x=0.25$.  At low $x$ the magnetic 
response is much weaker; note the difference in the scale for the upper two 
curves on Fig. \ref{magneti}.
This behavior was interpreted by Itoh et 
al. \cite {itoh} as evidence for a spin glass like phase.  

The high field magnetizations at 10K are presented in Table 1 along with the
average magnetizations per Co and per Sr atoms in units of $\mu_B$.
We found about 30\% higher magnetization than 
Se\~nar\'{\i}s-Rodr\'{\i}guez and Goodenough \cite{goodenough2} did for similar
compositions.  The high field magnetization per Co atom measured by Itoh 
{\it et al.} \cite{itoh} on the $x$=0.5 sample is also higher than that 
reported in ref \cite{goodenough2}.
The difference may be due to the different cooling rates of the 
samples. 

According to the data, the magnetization per Co atom increases approximately 
linearly with doping concentration.  Each Sr atom brings 5-7 spins to the 
system. The high value of magnetization per Sr site indicates that each 
dopant atom converts about two Co atoms into high (or intermediate) spin 
configuration.

\begin{figure}
\pswdincr=-1cm
\psboxto(3.4in;0in){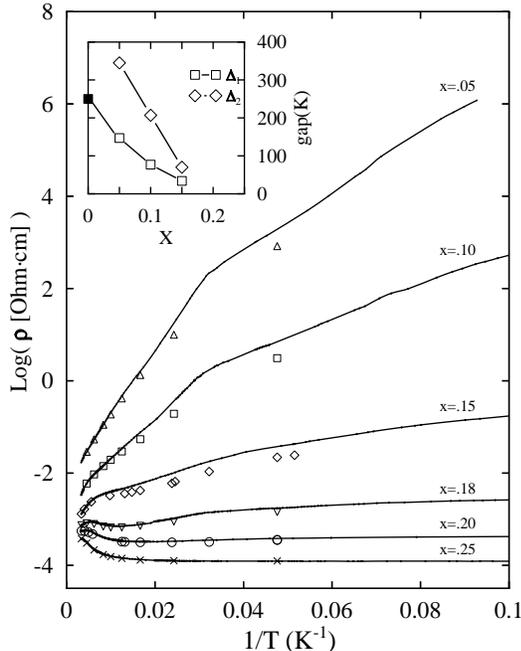}
\caption{Logarithm of resistivity {\it vs.} inverse temperature for a 
set of samples of various doping level $x$.  The solid line is the 
result of the measurement in zero magnetic field.  
Note the crossover 
between two distinct activation energies for low $x$ and 
the metallic behavior 
at high $x$.  The symbols are resistivities measured in H=7T.
The inset shows (open symbols) the activation energies 
evaluated from the slopes of resistivity curves for semiconducting samples.
Also shown in the inset (filled square) is the energy gap obtained for the $x$=0 sample 
from thermal expansion measurements by Asai {\em et al. }  
\protect\cite{asai}. The solid line in the inset is guide to the eye.
}
\label{resisti}
\end{figure}

In Figure \ref{resisti} the continuous lines represent the dc resistivity of the samples. 
The room temperature
resistivity of $x=0.2$ sample turned out to be higher than that of $x=0.18$ 
compound.  A systematic error, caused by geometrical factors, may
be responsible for this \cite{ganguly}. 
In the Figure the curve corresponding to $x=0.18$ was scaled 
up and the 
curve corresponding to $x=0.20$ was scaled down by a factor of 1.4. 
For low 
concentrations of Sr the samples  are semiconductors.  There are two 
distinct energy gaps in the semiconducting state: at higher temperatures 
($>30$K) the conductivity is characterized by a larger gap; at low 
temperatures ($<30$K)
a lower gap is observed.  The cross-over behavior is common for 
doped semiconductors \cite{textbook}; we will discuss this matter later.
For $x=0.18$ the conductivity shows the signs of 
a metal-insulator transition.  The resistivity of this sample drops 
dramatically at high temperature and approaches the metallic resistance of 
the highly doped samples.  The the magnitude and temperature dependence 
of the resistivity of the $x$=0.20 and 0.25 samples is metallic.  

\begin{figure}
\pswdincr=-3cm
\psboxto(3.4in;0in){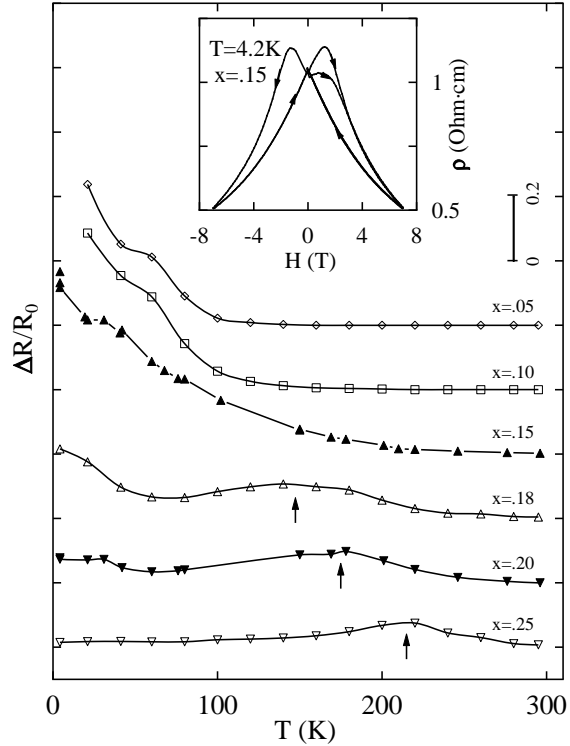}
\caption{Magnetoresistance, $\Delta R/R_0$, as a function of 
temperature.  The curves are shifted along the vertical axis for better 
view and the MR scale is indicated by the bar on the upper left side.  
The MR is close to zero at room temperature for all samples. The arrows indicate 
the ferromagnetic transition temperature. }
\label{magresi}
\end{figure}

In the Mn analogue of the material, the highest magnetoresistance has 
been observed in the neighborhood of the ferromagnetic transition.  
Figure \ref{magresi} illustrates that a similar behavior was found in our metallic 
samples:  there is an MR peak near the Curie temperature of the 
$x$=0.18-0.25 compounds.  However, we found even larger values of MR in 
the semiconducting phase, and the highest MR was observed in the low 
temperature spin glass regime. The magnetoresistance exhibits a hysteresis as 
it follows the internal magnetic fields in the sample, which lags behind 
the externally applied magnetic field (Fig. \ref{magresi}. inset).  
The resistivity in 7 T magnetic field, as 
obtained from field sweeps similar to that shown in the inset of the
Fig. \ref{magresi}., is represented in Fig. \ref{resisti} by empty symbols.    

In order to understand the electronic transport in the doped samples, we first 
consider the pure material, LaCoO$_3$.  The ground state electronic 
configuration of Co atom is $t_{2g}^6e_g^0$ with zero spin \cite{goodenough3}. 
The thermal excitation of Co atoms to the high spin $t_{2g}^4e_g^2$ 
(Co$^{3+}$) state
is responsible for the anomalous thermal expansion 
of pure LaCoO$_3$ \cite{asai}. The concentration 
$n$ of excited Co atoms can be estimated as \cite{asai} 
\begin{equation}
n={\nu \over \nu+ \exp (\Delta/k_BT) }
\label{concentration}
\end{equation}
where $\nu=15$ is the multiplicity of the high spin state.

Remarkably, the 
Co low spin $\rightarrow$ high spin transition gap in pure LaCoO$_3$, 
estimated from thermal expansion \cite{asai}, 
coincides very well with 
the activation energy determined from our low temperature resistivity 
measurements, if  the data are extrapolated to $x=0$ 
(Fig. \ref{resisti}, inset).  This coincidence suggests  
that the low temperature conduction is intimately related  to
the thermal activation of high spin states with a 
gap modified with doping, presumably due to lattice distortion. 
At high temperatures another activated process dominates the conduction, 
characterized by a conductivity of $\sigma_2 \exp (-\Delta_2/k_BT)$.  
The experimental data for $x=0.05$, 0.10 and 0.15 on Fig. \ref{resisti} are 
reasonably well fitted by the empirical formula
\begin{equation}
\rho^{-1}=\sigma_1 n'  + \sigma_2 \exp (-\Delta_2/k_BT)
\label{rhofit}
\end{equation}
where $n'= \nu /[ \nu+ \exp (\Delta_1/k_BT)]$  is the number of 
excited Co$^{3+}$ atoms.
The parameters $\sigma_1$, $\Delta_1$, $\sigma_2$, $\Delta_2$ are 
0.025$\Omega^{-1}$cm$^{-1}$ 140K, 150$\Omega^{-1}$cm$^{-1}$, 340K for x=0.05;
0.66$\Omega^{-1}$cm$^{-1}$ 90K, 400$\Omega^{-1}$cm$^{-1}$, 210K for x=0.10;
0.20$\Omega^{-1}$cm$^{-1}$ 40K, 500$\Omega^{-1}$cm$^{-1}$, 80K for x=0.15, 
respectively.  

In La$_{1-x}$Sr$_x$CoO$_3$ the dopant Sr introduces high spin ($t_{2g}^3e_g^2$) 
Co$^{4+}$ into the system \cite{chainani}.  An electron residing on the thermally 
excited Co$^{3+}$ can move to the Co$^{4+}$ sites {\it via} double exchange. 
At low temperature, when $n' \ll x$, the charge transport happens by hopping 
between Co$^{4+}$ sites with the number of carriers determined by the number
of excited Co$^{3+}$ atoms. The overlap
between the corresponding orbitals depends strongly on the doping level.  
This explains the small value and strong (exponential) $x$ dependence of the 
factor $\sigma_1$ in Eq. \ref{rhofit}. 

At temperatures above 30K the number of excited Co$^{3+}$ becomes 
greater than the dopant concentration $x$, therefore it is more 
appropriate to consider an array of Co$^{3+}$ sites to which Co$^{4+}$ donates
a hole that can jump, again using double exchange. The resulting band is nearly
full, and the number of holes is $x$, resulting in a 
$\sigma_2$ which scales approximately linearly with the doping.  
Disorder causes localization of the electronic states close to the 
band edge in the vicinity of the Fermi level.  Nevertheless, since nearly all
Co sites participate in the conduction, the overlap integral is large,
leading to high mobility carriers at an energy $\Delta_2$ below the 
Fermi energy. This is in accordance with $\sigma_2 \gg \sigma_1$.

The highest magnetoresistance was observed at low doping levels and low 
temperatures, where Itoh {\it et al.} \cite{itoh} suggested spin glass  
behavior.  The double exchange conduction in this state is strongly affected 
by the disorder in the spin distribution.  This disorder is partially 
suppressed by external magnetic field resulting in high, negative 
magnetoresistance.

In our metallic samples the MR is five times smaller than 
that observed in the spin-glass state, and it is also much smaller than 
the MR of the MnO$_3$ perovskites.
In a recent work, Millis {\it et al.} \cite{millis} argue that double 
exchange can not be the sole source of the anomalous large magnetoresistance 
in the  LaSrMnO compound. They suggest that the Jahn-Teller effect due to the 
displacement of oxygen around the Mn$^{3+}$ ion plays an important role.  
This mechanism sensitively depends on the presence of an unpaired 
electron on the upper $e_g$ level and therefore may not be active in 
our samples.  In accordance with the arguments of Millis {\it et al.}, the
significantly lower value of MR in Co compounds corresponds to the 
double-exchange mechanism alone.  

\bigskip

\widetext
\begin{table}
\caption{Magnetizations of La$_{1-x}$Sr$_x$CoO$_3$ for different
$x$ at 5 T and 10K}
\begin{tabular}{lcccccc}
$x$&0.05&0.10&0.15&0.18&0.20&0.25\\
magnetization, $emu \over{g}$&8.24&11.62&19.14&29.05&31.88&36.31\\
magnetic moment per Co atom, $\mu_B$&0.36&0.50&0.82&1.23&1.34&1.52\\
magnetic moment per Sr atom, $\mu_B$&7.2&5.0&5.4&6.8&6.7&6.1\\
\end{tabular}
\label{magnetization}
\end{table}
\narrowtext


The authors wish to thank J.M. Tranquada for initiating this study and for 
useful discussions,  and L. Henderson Lewis for help in sample preparation.
Work at SUNY, Stony Brook, is supported by the NSF grant DMR9321575.
Work at BNL is supported by the US Department of Energy, Division of 
Materials Science, under contract DE-AC02-76CH00016.

\end{document}